\documentclass[11pt,twoside]{article}

\usepackage{asp2006}
\usepackage{epsf}
\usepackage{lscape}

\begin{document}
\title{Jet/environment interactions in low-power radio galaxies}   
\author{J.H. Croston}   
\affil{Centre for Astrophysics Research, Science and Technology
  Research Institute, University of Hertfordshire, College Lane,
  Hatfield, Hertfordshire, AL10 9AB, UK.}    

\begin{abstract} 
The interactions between low-power radio galaxies and their
environments are thought to play a crucial role in supplying energy to
offset cooling in the centres of groups and clusters. Such
interactions are also important in determining large-scale radio
structures and radio-source dynamics. I will discuss new {\it
XMM-Newton} observations of the hot-gas environments of a
representative sample of nine FRI radio galaxies, which show strong
evidence for the importance of such interactions (including direct
evidence for heating) and provide important new constraints on source
dynamics and particle content. In particular I will show that the
widely discussed apparent imbalance between the internal lobe pressure
available from relativistic electrons and magnetic field and the
external pressure of hot gas correlates with radio structure, so that
naked jets require a large contribution from non-radiating particles,
whereas lobed sources do not. This may provide the first direct
observational evidence that entrainment of the ICM supplies the
missing pressure.

\end{abstract}

\section{Introduction}

Low-power (FRI) radio galaxies are common in groups and clusters of
galaxies, where they are thought to play an important role as part of
feedback processes regulating gas cooling and galaxy evolution. The
dynamics and evolution of the jets in FRI radio galaxies are
essentially controlled by interactions with their hot-gas environments
on galaxy, group and cluster scales, resulting in a wide variety of
large-scale radio structures ranging from narrow tails to rounded
lobes. The environmental impact of an FRI radio galaxy is highly
dependent on its large-scale morphology, so that it is important to
understand in which environmental conditions different types of
structures are produced. {\it ROSAT} observations first established
that the brightest and most well-studied FRI radio galaxies typically
reside in group-scale hot-gas environments \citep[e.g.][]{kom99,wb00},
while the majority of cool-core clusters also possess low-power
sources (often of a more amorphous morphology). With {\it Chandra} and
{\it XMM-Newton} it is possible to study radio-galaxy environments,
and jet/environment interactions in more detail
\citep[e.g.][]{c03,bla01,fab06}; however, to date the environmental
properties of only a few of the nearby, well-studied FRI radio
galaxies have been studied in depth.

One of the key uncertainties in modelling radio-galaxy dynamics and in
establishing their energetic impact is the lack of constraints on
their particle content. In the case of powerful (FRII) radio galaxies,
two lines of evidence suggest that synchrotron minimum energy
estimates are fairly reliable: (1) measurements of inverse-Compton
emission from the lobes of FRII radio galaxies suggest that their
electron energy densities are close to the value for equipartition in
the absence of an energetically important proton population
\citep[e.g.][]{c05a,ks05}, which would have to be coincidental if such
a proton population was present; (2) for cases where both external
pressure measurements and inverse-Compton internal electron pressures
are available, they are in reasonable agreement
\citep[e.g.][]{mjh02,c04,bel04} so that there is no requirement for an
additional particle population on energetic grounds. However, in
low-power (FRI) radio galaxies, it has been known for several decades
that the radio-emitting relativistic electrons, together with an
equipartition magnetic field, cannot provide sufficient pressure to
balance the measured external pressure if it is at equipartion
\citep[e.g.][]{mor88,har98,wb00}. In \citet{c03}, we showed for two
FRIs that the additional pressure cannot be due solely to electron
dominance as this would result in detectable X-ray inverse-Compton
emission, which is not seen, and the presence of sufficient thermal
gas at the temperature of the environment was also ruled out by the
presence of deficits in X-ray surface brightness at the positions of
radio lobes. These conclusions also apply for the cluster
cavity systems studied by \citet{dun04} and \citet{bir04}. The most
plausible origin of the required additional pressure is therefore
either material that has been entrained and heated, or magnetic
pressure; however, these scenarios are both difficult to test.

Here I present preliminary results of an {\it XMM-Newton} study of a
representative sample of nine low-power (FRI) radio-galaxy
environments, with the aims of constraining particle content and
measuring the environmental impact of FRI radio galaxies with a wide
range of large-scale morphologies. The sample consists of 3C\,76.1,
NGC\,1044, 3C,296, 3C\,31, NGC\,315 (new {\it XMM-Newton}
observations), 3C\,66B, 3C\,449, NGC\,6251 (observations published in
\citealt{c03} and \citealt{ev05}), NGC\,4261 (archival data
set). All nine radio galaxies are found to lie in group-scale hot-gas
environments, with bolometric X-ray luminosities ranging from
$\sim 10^{41}$ -- $10^{43}$ erg s$^{-1}$. Full details of the X-ray
analysis will be published in Croston et al. (in prep).

\section{Relationship between radio and X-ray structure}
\label{xraydio}
\begin{figure*}
\epsfxsize 7.0cm
\epsfbox{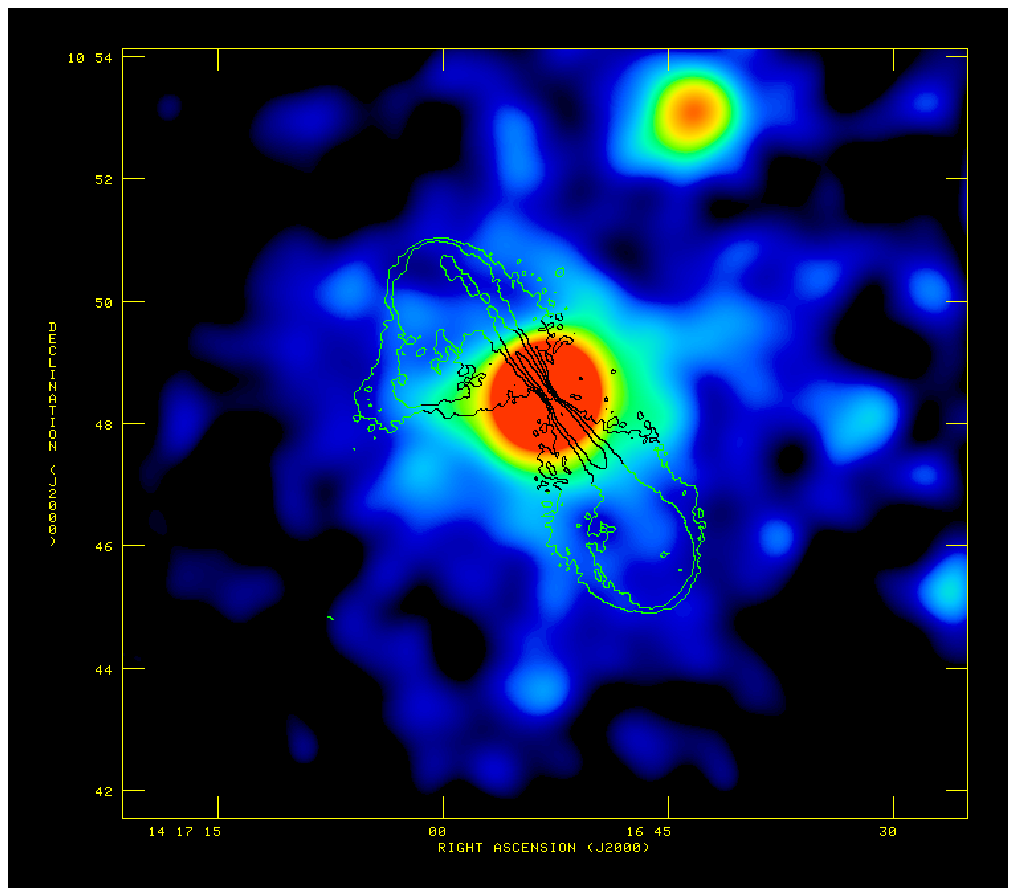}
\hskip 10pt
\epsfxsize 6.4cm
\epsfbox{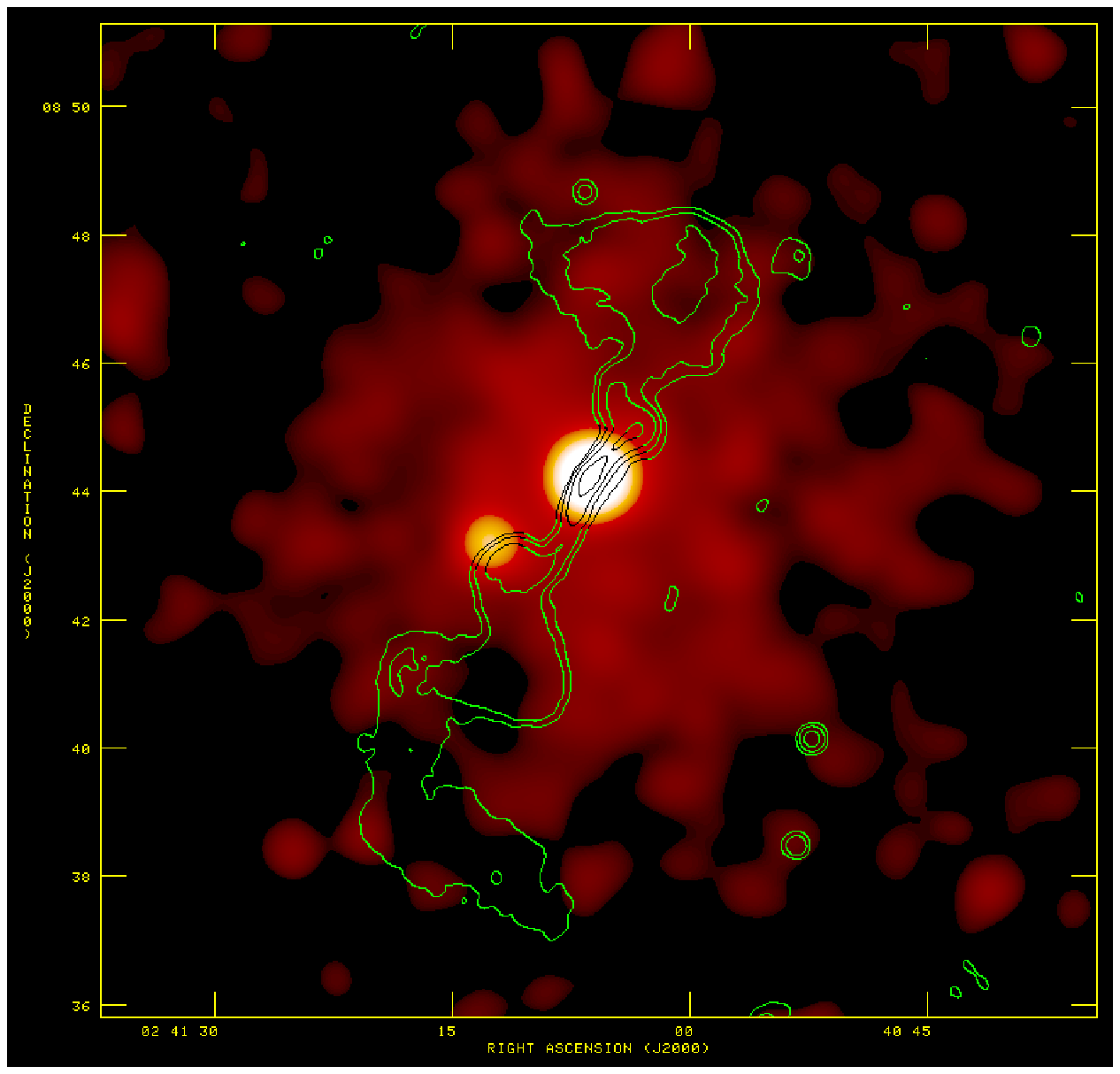}
\caption{Combined {\it XMM-Newton} MOS1, MOS2 + pn image of the X-ray
  environments of 3C\,296 (left) and NGC\,1044 (right) with radio
  contours overlaid, indicating X-ray cavities at the positions of the
  radio lobes.}
\label{cavs}
\end{figure*}

The importance of jet/environment interactions for group and cluster
gas has been highlighted by the large number of X-ray cavity systems
now known, ranging from elliptical galaxies to the richest clusters
\citep[e.g.][]{dun04, bir04}. Of the nine systems studied here with
{\it XMM-Newton}, 6 have clear, statistically significant detections
of cavities associated with the radio lobes (two new examples are
shown in Fig~\ref{cavs}). In several cases where a cavity is not
identifiable in the X-ray data, the emission is nevertheless elongated
perpendicular to the direction of the radio axis, thus demonstrating a
clear link between the structure of the X-ray emitting gas and the
radio source. While the distribution of group gas clearly indicates an
important relationship between the radio and X-ray properties, I found
no correlations between global X-ray and radio properties, including
luminosities, sizes, and structural parameters. It is unclear which
properties control the radio luminosity and power of the group-scale
AGN outbursts, but this work suggests that small-scale properties
related to AGN fuelling may play a stronger role than large-scale
environment in determining the luminosity and size of the radio
structure.

\section{Particle content}

\begin{figure}[!b]
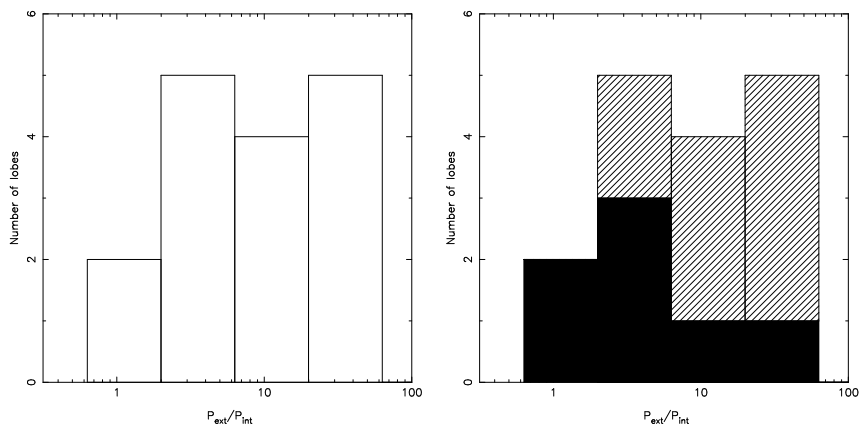

\begin{center}
\epsfxsize 5.5cm
\epsfbox{f2a.ps}
\hskip 5pt
\epsfxsize 5.5cm
\epsfbox{f2b.ps}
\caption{Histogram of the ratios of external to internal (minimum)
  radio-lobe pressure for the full sample of 9 radio galaxies (16
  lobes), right; separated into `naked jet' sources (filled) and `lobed' sources
  (hatched), right.}
\label{pressures}
\end{center}
\end{figure}

In agreement with previous studies using earlier X-ray observatories
\citep[e.g.][]{mor88,wb00} and using {\it XMM-Newton}
\citep[e.g.][]{c03}, we found that the measured external pressures
from the hot gas environments were in nearly all cases significantly
higher than the minimum internal pressures for the radio lobes.
Fig.~\ref{pressures} (left) shows a histogram of the pressure ratios
($P_{ext}/P_{int}$) for the 16 lobes in the sample (only one lobe is
within the {\it XMM-Newton} field-of-view for NGC\,315 and NGC\,6251)
showing that the lobes are underpressured by factors ranging from
$\sim 0.02$ to 1.

\begin{figure*}[!b]
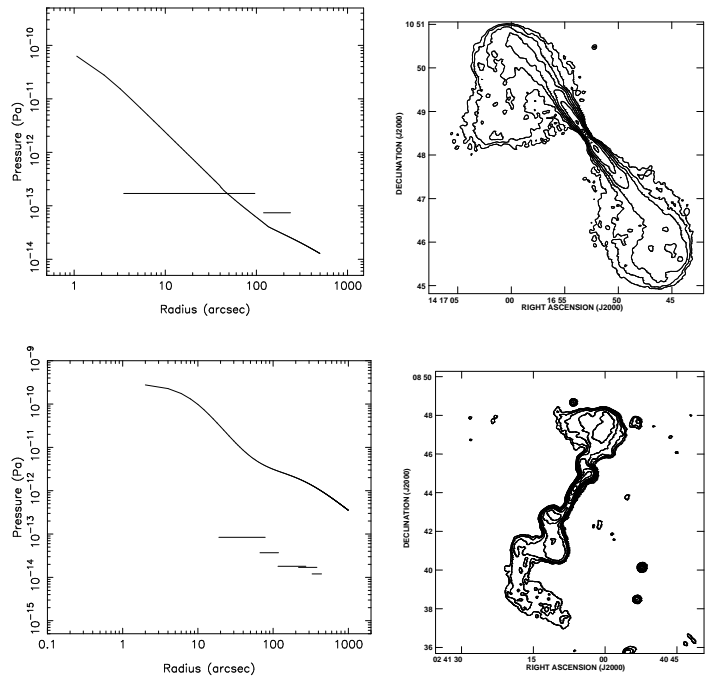

\begin{center}
\epsfxsize 4.7cm
\epsfbox{f3a.ps}
\hskip 5pt
\epsfxsize 4.3cm
\epsfbox{f3b.ps}
\vskip 10pt
\epsfxsize 4.8cm
\epsfbox{f3c.ps}
\hskip 5pt
\epsfxsize 4.2cm
\epsfbox{f3d.ps}
\caption{External pressure profiles for 3C\,296 (top left) and
  NGC\,1044 (bottom left) with horizontal lines indicating internal
  minimum lobe pressures for different components of the source. The
  corresponding radio maps are shown to the right of each profile to
  illustrate the apparent link between radio morphology and pressure
  deficit as shown in the right-hand panel of Fig.~\ref{pressures}.}
\label{profs}
\end{center}
\end{figure*}

This is the first time that a study has been carried out for a
reasonable sized sample that includes the full range of observed FRI
radio morphologies, and so a primary aim was to investigate whether
the amount of ``missing'' pressure is related to the properties of the
radio-source and/or its environmental interaction. Fig.~\ref{profs}
shows the external pressure profiles for the two sources with the most
extreme behaviour: 3C\,296 (top left) appears to be in approximate
pressure balance, assuming equipartition, so that no additional
pressure contribution is required; NGC\,1044 (bottom left) has a high
ratio of $P_{ext}/P_{int}$ ($\sim 40$). Fig.~\ref{profs} also shows
the corresponding radio maps for the two sources. It is clear that
their radio morphologies are quite different: NGC\,1044 has naked jets
that remain fairly well collimated over large distances before
gradually spreading into plumes or tails, whereas the jets of 3C\,296
appear to enter radio lobes at a small distance from the nucleus. This
morphological difference suggests a model in which jet/environment
interactions affect the particle and/or energy content of the radio
lobes; for example, the dominant factor in determining the ratio of
radiating to non-radiating particles may be the efficiency of the
entrainment process.

To test this hypothesis, I separated the 16 radio lobes into the two
categories of `naked jet' sources and `lobed' sources. The sources
classified as having `naked jets' are: 3C\,31, NGC\,1044, NGC\,315,
3C\,449, NGC\,6251; those categorised as `lobed' are: 3C\,296,
3C\,76.1, NGC\,4261, 3C\,66B. The right-hand panel of
Fig.~\ref{pressures} shows a histogram of the pressure ratios for the
two categories, which demonstrates that there is indeed a signficant
difference in the two populations in the expected sense: the `lobed'
sources are typically closer to pressure balance at equipartition than
the `naked jet' sources. This comparison therefore supports the
hypothesis that entrainment of the jet's surroundings on scales of
tens to hundreds of kpc is responsible for the apparent pressure
imbalance of FRI radio galaxies. Unfortunately there is no obvious
observational proxy for jet deceleration or entrainment that can be
used to test this hypothesis further with the current data.
Comparisons with the models of \citet[][and in prep.]{lb02} show that
naked jet sources such as 3C\,31 have an entrainment rate that
increases with distance from the nucleus, whereas lobed sources such
as 3C\,296 may not.

If this explanation is correct for the FRIs studied here, then an
obvious question is whether it is consistent with the pressure offsets
observed in cluster centre radio sources \citep[e.g.][and McNamara et
al., this meeting]{dun04,bir04}, which typically do not possess
well-collimated jets on large scales. Given the amorphous structures
of many cluster centre sources, suggestive of strong environmental
influence, it seems plausible that there is considerable mixing of
cluster gas with radio-lobe plasma (and heating, as required in order
for the cavities to be detected). In addition, \citet{dun06} argue,
based on a comparison of the leptonic content of the small-scale jet
and the energetics of the cluster bubbles, that even in bubbles with
large apparent pressure offsets such as those of the Perseus cluster,
the jet is leptonic, so that the pressure offset on large scales is
more likely to be due to entrainment rather than a relativistic proton
population, consistent with the picture I present here. Hence, while
the categorization of group-scale sources into `lobed' and `naked-jet'
morphologies with different energetics may not apply for cluster
centre sources, these results may nevertheless provide some insight
into the solution of the pressure balance problem for these systems as
well.

\section{Environmental impact}

As discussed in Section~\ref{xraydio} and shown in Fig.~\ref{cavs}, it
is clear that the group gas surrounding FRI radio galaxies is strongly
affected by their presence. Earlier {\it XMM-Newton} studies of the
group environments of FRIs found direct evidence for localised heating
of the group gas, and indirect evidence for more generalised heating
\citep{c03}. In \citet{c05b} I reported an apparent systematic
difference in the radio properties of galaxy groups with and without
central radio sources, in the sense that radio-loud groups appear to
be systematically hotter for a given luminosity. A recent study
comparing REFLEX/NORAS clusters with NVSS has found similar results
\citep{mag07}, and a recent {\it Chandra} study of a subset of the
\citet{c05b} sample by \citet{jet07} has shown that the gas properties
in the innermost regions of the group do not appear to be
significantly affected by the presence of a radio source, so that any
heating processes must be most significant on larger scales.

\begin{figure}[!t]
\begin{center}
\epsfxsize 6.5cm
\epsfbox{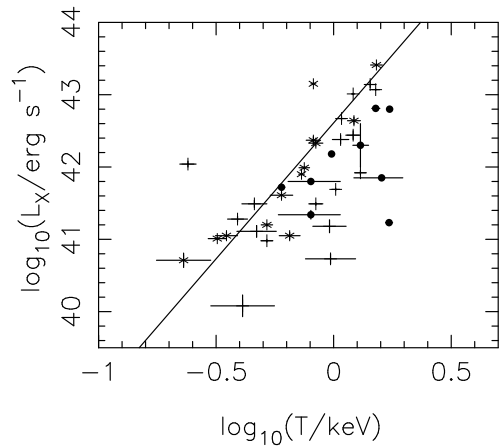}
\caption{The $L_{X}/T_{X}$ relation for the {\it
    XMM-Newton} radio-galaxy environments sample (filled circles)
    compared with the radio-quiet (stars) and radio-loud (+ symbols)
    samples of Croston et al. (2005). Solid line indicates the
    best-fitting relation for radio-quiet groups.}
\label{lt}
\end{center}
\end{figure}

In Fig.~\ref{lt} I show the X-ray luminosity-temperature relation for
the current sample of 9 radio galaxy group environments (filled
circles). The $\times$ and $+$ symbols indicate the radio-quiet and
radio-loud samples of \citet{c05b}, respectively, and the solid line
indicates the best-fitting relation for the radio-quiet groups. The
{\it XMM-Newton} radio-galaxy sample all lie to the high temperature
side of the radio-quiet relation, thus confirming the earlier result:
groups containing extended radio sources scatter to higher
temperatures for a given X-ray luminosity. In \citet{c05b}, we argued,
based on a lack of difference in the X-ray-to-optical luminosity
ratios for the two subsamples, that this effect must be due to a
temperature increase, rather than a luminosity decrease, as might be
expected. The origin of this effect therefore remains unclear: for the
current sample, the $P$d$V$ work available from the currently
observable radio source is in most cases insufficient to provide the
energy to heat the cluster gas by the required amount, and there is no
correlation between radio luminosity and temperature excess.

\section{Summary}
Study of a representative sample of nine low-power (FRI) radio-galaxy
environments with {\it XMM-Newton} has led to the following
conclusions:
\begin{itemize}
\item Low-power radio galaxies typically inhabit group-scale
  environments, which can range in X-ray luminosity from $10^{41} -
  10^{43}$ erg s$^{-1}$. It is therefore important for models of
  radio-galaxy evolution to take into account the wide range of
  possible environments in which these sources can occur.
\item The apparent pressure imbalance seen in low-power radio galaxies
  appears to be related to radio-source morphology. `Naked jet'
  sources require a large contribution from non-radiating particles or
  magnetic pressure, whereas `lobed' sources do not. This suggests
  that entrainment efficiency may be the dominant factor in
  determining the energetics of the large-scale radio structure in
  FRIs.
\item The environments of all nine radio-galaxies in this sample
  appear to be hotter than is predicted by the $L_{X}/T_{X}$ relation
  for radio-quiet groups. The origin of this apparent heating effect
  remains unclear.
\end{itemize}


\acknowledgements 

I would like to thank the following people who have contributed to my
work on the environments of low-power radio galaxies: Martin
Hardcastle, Mark Birkinshaw, Diana Worrall, Robert Laing and Dan
Evans.


\end{document}